\documentclass[10pt, twocolumn]{IEEEtran}
\usepackage{amsmath}
\usepackage{amssymb}
\usepackage{mathrsfs}
\usepackage{cite}
\usepackage{epsfig}
\usepackage{epsf}
\usepackage{theorem}
\usepackage{graphics}
\usepackage[active]{srcltx}

\begin{document}

\title{Energy Efficiency and Reliability in \\
Wireless Biomedical Implant Systems
\thanks{Manuscript received November 28, 2010; accepted January 4, 2011.
The work was supported in part by an Ontario Research Fund (ORF) project
entitled ``Self-Powered Sensor Networks''. The work of J. Abouei was
performed when he was with the Department of Electrical and Computer
Engineering, University of Toronto, Toronto, ON M5S 3G4, Canada.}
\thanks{J. Abouei is with the Department of Electrical and Computer Engineering,
Yazd University, Yazd, Iran (e-mail: abouei@yazduni.ac.ir). }
\thanks{J. D. Brown is with the DRDC Ottawa, Ottawa, ON, Canada. He completed his
Ph.D. in ECE at University of Toronto (e-mail:
david\_jw\_brown@yahoo.com).}
\thanks{K. N. Plataniotis, and S. Pasupathy are with the Edward S.
Rogers Sr. Department of Electrical and Computer Engineering,
University of Toronto, Toronto, ON M5S 3G4, Canada (e-mails:
\{kostas, pas\}@comm.utoronto.ca).}}

\author{\small Jamshid Abouei, \emph{Member, IEEE,} J. David
Brown, Konstantinos N. Plataniotis, \emph{Senior Member, IEEE} and \\
Subbarayan Pasupathy, \emph{Life Fellow, IEEE}}

\maketitle

\markboth{IEEE TRANSACTIONS ON INFORMATION TECHNOLOGY IN BIOMEDICINE}{}

\begin{abstract}
The use of wireless implant technology requires correct delivery of the
vital physiological signs of the patient along with the energy management in
power-constrained devices. Toward these goals, we present an
augmentation protocol for the physical layer of the Medical Implant
Communications Service (MICS) with focus on the energy
efficiency of deployed devices over the MICS frequency band. The
present protocol uses the rateless code with the Frequency Shift Keying
(FSK) modulation scheme to overcome the reliability and power cost
concerns in tiny implantable sensors due to the considerable
attenuation of propagated signals across the human body. In
addition, the protocol allows a fast start-up time for the
transceiver circuitry. The main advantage of using rateless codes is to
provide an inherent adaptive duty-cycling for power management, due
to the flexibility of the rateless code rate. Analytical results
demonstrate that an 80\% energy saving is achievable with the
proposed protocol when compared to the IEEE 802.15.4 physical layer
standard with the same structure used for wireless sensor networks.
Numerical results show that the optimized rateless coded FSK is more
energy efficient than that of the uncoded FSK scheme for deep tissue
(e.g., digestive endoscopy) applications, where the optimization is
performed over modulation and coding parameters.

\end{abstract}


\section{Introduction}
\subsection{Background}
Recent advances in wireless sensor technologies have opened up a new
generation of ubiquitous body-centric systems, namely Wireless Body
Area Networks (WBANs), for providing efficient health care services
\cite{Shmidt2002}. WBANs support a broad range of
medical/non-medical applications in home/health care, medicine,
sports, etc. Of interest is the use of WBANs for the continuous
remote monitoring of the vital physiological signs of patients, regardless
of the constraints on their locations and activities
\cite{Varshney_ICM0406}.

\begin{table}
\label{table001} \caption{Comparing some MICS specifications with the IEEE
802.15.4 Standard used for WSNs} \centering
  \begin{tabular}{lcc}
  \hline
  ~~~~\textbf{Parameter}              &  \textbf{MICS}                   &  \textbf{IEEE 802.15.4}             \\
  \hline
  Frequency Band        & $402-405$ MHz           & $868/915$ MHz (Eur./US)     \\
                        &                         & $2.4$ GHz (Worldwide)      \\
  Bandwidth             & $300$ KHz               & $62.5$ KHz (for $2.4$ GHz) \\
  Data Rate             & support more than              & $250$ kbps (for $2.4$ GHz) \\
                        & $250$ kbps                     &  \\
  Transmit Power        & $25$ $\mu$w ($-16$ dBm) &  $0.5$ mw ($-3$ dBm)                  \\
  Operating Range       & Typically $0-2$ m       & Typically $0-10$ m         \\
  \hline
  \end{tabular}
\end{table}

Broadly, WBANs can be classified into two categories of wearable
(on-body) and implantable (in-body) systems \cite{ReusensITITB1109,
YazdandoostIEICE0209}. A wearable WBAN provides RF communications
between on-body sensors and a central controller for the remote
monitoring of vital signs of a patient such as ECG, blood pressure
and temperature. Such a wearable structure is a short-range communication system
based on the Wireless Medical Telemetry Service (WMTS)
officially adopted by the Federal Communications Commission (FCC)
\cite{FCC2000}. On the other hand, in an implantable WBAN, a
biosensor embedded within the body communicates with an
apparatus sticking on the body or with an external monitoring device
in the vicinity of the human body. Examples of implantable devices
are cardiac pacemakers, insulin dispensers, neurostimulators and
bladder controllers \cite{Receveur2007}. Such a wireless implantable
technology is a very short-range communication system used for monitoring health
conditions of patients such as brain waves in paralyzed persons,
glucose levels in diabetic patients and patients with gastrointestinal disorders.
The medical implantable devices operate according to the Medical
Implant Communications Service (MICS) rules established by FCC
\cite{FCC2003}.

The MICS standard is distinguished from the existing
standards used for Wireless Sensor Networks (WSNs)
\cite{IEEE_802_15_4_2006, FreeScale2007} due to the size, power
consumption and frequency band requirements for implantable devices
(see Table I). For instance, the MICS protocol uses an 25 $\mu$W
(-16 dBm) Effective Isotropic Radiated Power (EIRP) at the 402-405
MHz frequency band. This provides a low power transmitter with a
small size antenna resulting in a compact and lightweight
implantable device, while, today's low power radio devices which use
standards such as ZigBee and Bluetooth (IEEE
802.15.1) cannot meet these stringent requirements. Furthermore,
the limit -16 dBm EIRP in the MICS standard reduces the effect of
the interference on the other biosensor devices operating in the
same frequency band \cite{FCC2003}. In addition, deploying
ultra low power devices minimizes the heat absorption and the
temperature increase caused by implantable biosensors, reducing the
thermal effects on the body tissues \cite{TangITBE0705}.

Besides the simplicity and low power consumption characteristics,
implantable devices must be robust enough to provide the desired
service for long-term applications. Furthermore, since biosensors
frequently switch between the sleep mode and the active mode, they should have
fast start-up times. The protocol used for
such implantable devices must support a high degree of reliability
and energy efficiency in a realistic channel model for the
human body. Each of these characteristics presents considerable
design challenges for different layers of implantable WBANs,
including the MAC layer and the physical layer
\cite{ Oh_ITITB2010, BohorquezIJSC0409, ChoITMT1009}. Central to the
study of the physical layer design of implantable WBANs is the
modulation and coding. Deploying proper modulation schemes which
minimize the total energy consumption in both circuit and RF signal
transmission prolongs the biosensors lifetime. Enhanced reliability,
on the other hand, is achieved by adding redundant information bits
in the data packets in the Forward Error Correction (FEC) coding
stage at the cost of an extra power consumption. Thus, there is
generally a trade-off between a higher reliability and a lower power
consumption in using FEC codes.

\subsection{Related Work and Contribution}
There are several works in the literature that study the physical
layer design of implantable WBANs. Oh \emph{et al.} \cite{Oh_ITITB2010}
propose a new type of phase shift keying modulation scheme, namely
Phase Silence Shift Keying (PSSK), for high data rate implantable
medical devices which is more bandwidth efficient than orthogonal
modulation schemes such as OOK. Under an Additive White Gaussian
Noise (AWGN) channel model with path loss for the human body, reference
\cite{Oh_ITITB2010} shows that the Bit Error Rate (BER) of an 8-PSSK
is lower than that of various sinusoidal carrier-based modulation
schemes. Reference \cite{BohorquezIJSC0409} proposes a simple and
ultra low power hardware design for an FSK/MSK direct modulation
transmitter in medical implant communications which supports the
data rate and the EIRP requirements in the MICS standard. However,
no channel coding scheme and energy efficiency analysis were
considered in \cite{Oh_ITITB2010} and \cite{BohorquezIJSC0409}.
Reference \cite{YuceITMT1009} investigates the feasibility of
applying an Ultra-Wideband (UWB) scheme to implantable applications.
However, according to the FCC regulations in \cite{Oppermann2004},
UWB technologies with a typically
3-10 GHz frequency band are used for transmitting information over a
wide bandwidth at least 500 MHz, which are wider than the frequency
band and the bandwidth used in the MICS standard.

Since, the human body is a lossy medium, and due to the adverse
circumstances in the patient's body (e.g., variable body temperature
due to the fever, changing the location of the implant capsule in
the digestive system), the RF signal transmitted from the in-body
biosensor experiences a dynamic channel environment to reach to its
corresponding on-body receiver. In addition, it is shown in
\cite{SaniITMT1009} that the efficiency of wireless implantable
devices is related to the human body structure (e.g., male
or female, body size, obese or thin and tissue composition).
Clearly, finding a proper FEC coding scheme which overcomes the
patient's vital information loss, and operates in a high degree of
flexibility and accuracy over the above dynamic channel conditions
is desirable. Existing protocols use various methods to overcome the
packet loss and decoding error concerns in classical wireless sensor
networks (e.g., see \cite{VuranIACM0409}). The Automatic Repeat
Request (ARQ) approach, for instance, requires many retransmissions
in the case of poor channel conditions, resulting in a high latency
in the network. This is in contrast to the transmission of patient's
real time vital signs which are sensitive to the latency.
More recently, the attention of researchers
has been directed toward deploying rateless codes (e.g., Luby
Transform (LT) code \cite{LubyFOCS2002}) in conventional wireless networks due to the
significant advantages of these codes in erasure channels.
However, to the best of our knowledge, there is no existing analysis
on the energy efficiency and reliability of rateless coded
modulation in implant WBANs.

This work deals with the first in-depth analysis on the energy
efficiency and reliability of a wireless biomedical implant system
which uses LT codes with the Frequency Shift Keying (FSK) modulation
scheme. The present analysis is based on a realistic channel model
for the human body as well as the requirements stated in the MICS
standard. The main outcome of this work is to introduce an
augmentation protocol for the physical layer of the MICS standard
for implant WBANs. The protocol allows a fast start-up time for
implantable devices along with the lower power consumption during
active mode duration. It is demonstrated that an 80\% energy saving
is achievable compared to the IEEE 802.15.4 physical layer protocol
with the same structure used for wireless sensor networks.
Furthermore, we show that an inherent adaptive duty-cycling for
power management is provided which comes from the flexibility of the
LT code rate. Numerical results for deep
tissue applications show a lower energy consumption in
the optimized coded scheme compared to the uncoded case.

The rest of the paper is organized as follows. In Sections
\ref{System_model} and \ref{Channel_Model}, the implant wireless
system based on  a realistic channel model for the human body is
described. The energy consumption of an uncoded FSK modulation
scheme is analyzed in Section \ref{uncoded_MFSK}. Design of LT codes
and the energy efficiency of the LT coded FSK are presented in
Section \ref{analysis_Ch4}. Section \ref{simulation_Ch5} provides
some numerical evaluations using realistic parameters to confirm our
analysis. Also, some design guidelines for using LT codes in
practical implant WBANs are presented. Finally in Section
\ref{conclusion_Ch6}, an overview of the results and conclusions are
presented.

\section{System Model and Assumptions}\label{System_model}
\begin{figure}[t]
\centerline{\psfig{figure=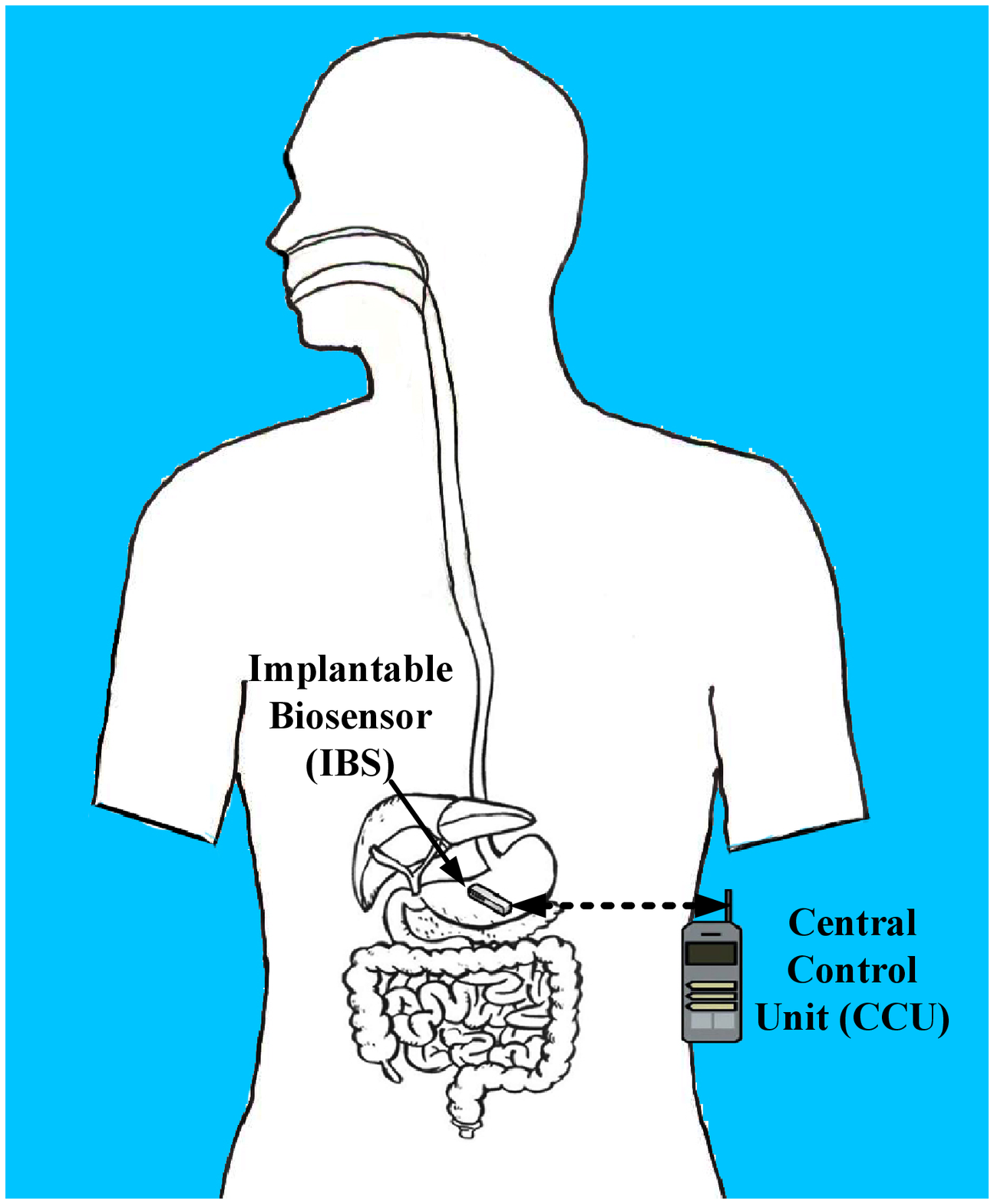,width=2.3in}} \caption{A
wireless biomedical implant system. } \label{fig: Implant-WBAN}
\end{figure}

In this work, we consider a wireless biomedical implant system
depicted in Fig. \ref{fig: Implant-WBAN} with a bidirectional
communication between an implantable biosensor, denoted by IBS, and
an external Central Control Unit (CCU), where the transmission in
each direction takes place in a half-duplex mode. We define the link
used for the transmission of signals from the IBS to the CCU as an
\emph{uplink}, while the link from the CCU to the IBS is represented
as a \emph{downlink}. During the uplink transmission period, the IBS
\begin{figure*}[t]
\centerline{\psfig{figure=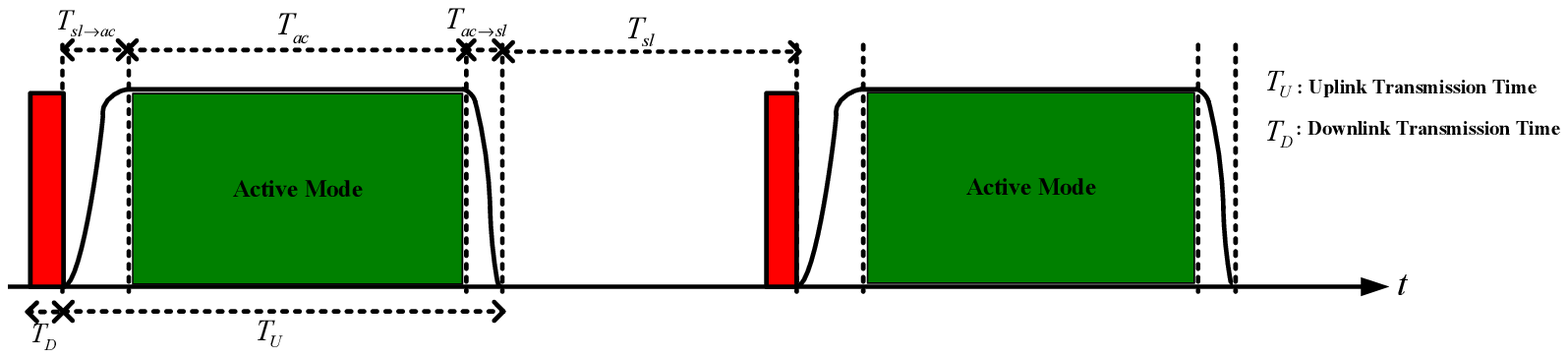,width=6.85in}}
\caption{A practical multi-mode operation in a reactive implant
WBAN. } \label{fig: Time-Basis}
\end{figure*}
wakes up after receiving a command signal from the CCU, and
transmits the processed raw physiological signals to the CCU. For
this period, the CCU is set to the listen mode. For the
downlink transmission interval, the IBS is switched to the listen
mode to receive control commands from the CCU. It is worth
mentioning that uplink has much more data to transmit than downlink,
meaning the downlink transmission time is much smaller than that of
the uplink. The main advantage of this half-duplex operation is that
a common carrier is shared between the uplink and the downlink (by
switching the bandwidth resource in time), resulting in further reduction
of the IBS complexity. In addition, in contrast to the full-duplex
systems, where they use two different frequency bands for
transmitting and receiving signals, using the half-duplex solution
maximizes the data rate for each link over the whole bandwidth. This
arises from the Shannon capacity formula
$C=B\log_2(1+\textrm{SNR})$, where the channel capacity $C$ is
proportional to the bandwidth $B$. The assumption of half-duplex
operation is used in many studies on this topic (e.g., see
\cite{SaniITMT1009} for implant WBANs). Longer transmission time for
the uplink compared to the downlink time interval makes the uplink
to be considered as a critical path in wireless in-body communications
from the power consumption point of view. More precisely, unlike
the CCU which has a power source, the IBS has power constraint, and
it needs to reduce its power consumption for minimizing the tissue
heating, extending the battery life in the IBS and meeting the power
restriction in the MICS standard. For this purpose, we introduce a
synchronized switching process which achieves a significant energy
saving in the uplink mode.

For the proposed communication system, the IBS and the CCU must
synchronize with one another and operate in a real time based
process known as \emph{duty-cycling} as depicted in Fig \ref{fig:
Time-Basis}. During the \emph{active mode} period $T_{ac}$ for the
uplink data path, the weak raw signal sensed by the implantable
biosensor is first passed through the amplification and filtering
processes to increase the signal strength and to remove unwanted
signals and noise. The filtered analog signal is then digitized by
an Analog-to-Digital Converter (ADC), and an $L$-bit binary message
sequence $\mathbb{M}_L = \lbrace{ m_i \rbrace}_{i=1}^{L}$ is
generated, where $L$ is assumed to be fixed. The bit stream is then
sent to the FEC coding unit. The encoding process begins by dividing
the message $\mathbb{M}_L$ into blocks of equal length denoted by
$\mathbb{B}_j = (m_{_{(j-1)k+1}},...,m_{jk})$,
$j=1,...,\frac{L}{k}$, where $k$ is the length of any particular
$\mathbb{B}_j$. Each block $\mathbb{B}_j$ is encoded by a rateless code
(in a manner to be described in Section \ref{analysis_Ch4}) to
generate a coded bit stream $\mathbb{C}_j =
(a_{_{(j-1)n+1}},...,a_{jn})$, $j=1,...,\frac{L}{k}$, with block
length $n$, where $n$ is a random variable and is determined based
on the channel condition.

The coded stream is then modulated by an FSK modulation scheme and
transmitted to the CCU across tissues in the human body. We will
explain later the advantages of using FSK modulation scheme in the
proposed implant system. Finally, the IBS
returns to the \emph{sleep mode}, and all the circuits of the
transceiver are powered off for the sleep mode duration $T_{sl}$ for
energy saving. It should be noted that only the wake-up circuit in
the IBS remains on during $T_{sl}$, and waits for a wake-up command
from the CCU. Since the wake-up circuit is battery-free (similar to
that used in passive Radio Frequency Identification (RFID) tag
technologies), its power consumption is very low
\cite{IEEE_P802_15}. We denote $T_{tr}$ as the \emph{transient mode}
duration consisting of the switching time from sleep mode to active
mode (i.e., $T_{sl \rightarrow ac}$) plus the switching time from
active mode to sleep mode (i.e., $T_{ac \rightarrow sl}$), where
$T_{ac \rightarrow sl}$ is short enough compared to $T_{sl
\rightarrow ac}$ to be negligible. Furthermore, the amount of power
consumed for starting up the IBS is more than the power consumption
during $T_{ac \rightarrow sl}$. Under the above considerations, the
IBS/CCU devices have to process one entire $L$-bit message
$\mathbb{M}_L$ during $0 \leq T_{ac} \leq T_L-T_{sl}-T_{tr}$ in the uplink
mode before a new sensed packet arrives, where $T_L =
T_{tr}+T_{ac}+T_{sl}$ with $T_{tr} \approx T_{sl \rightarrow ac}$.

Since an implant WBAN communicates over a very short-range across
the human body, the circuit power consumption is comparable to the
RF transmit power consumption. We denote the total circuit power
consumption in the uplink path as $P_c = P_{ct}+P_{cr}$,
where $P_{ct}$ and $P_{cr}$ represent the power consumption of the
circuits embedded in the IBS (as the transmitter) and the CCU (as
the receiver), respectively. In addition, the power consumption of
RF signal transmission in the IBS is denoted by $P_t$. Taking these
into account, the total energy consumption during the active mode
period in uplink, denoted by $E_{ac}$, is given by
$E_{ac}=(P_c+P_t)T_{ac}$. Since the downlink data path is only used
for transmission of command signals, we assume that the total energy
consumption (in both RF transmission and circuits) during the
downlink transmission period is negligible compared to that of the
uplink case. The energy consumption in the sleep mode period is
given by $E_{sl}=P_{sl}T_{sl}$, where $P_{sl}$ is the corresponding
power consumption. The present state-of-the art in implantable devices aims to keep a
very low sleep mode leakage current (coming from the CMOS circuits in the IBS),
resulting in $P_{sl}\approx 0$ when compared to the power consumption in the active
mode duration. This assumption is used in many
studies on the energy efficiency of wireless sensor networks (e.g.,
see \cite{Cui_GoldsmithITWC0905, Jamshid_TSP2011}).
As a result, the \emph{energy
efficiency}, referred to as the performance metric of the proposed
implant WBAN, can be measured by the total energy consumption during
the uplink transmission interval and corresponding to $L$-bit
message $\mathbb{M}_L$ as follows:
\begin{equation}\label{total_energy1}
E_L = (P_c+P_t)T_{ac}+P_{tr}T_{tr},
\end{equation}
where $P_{tr}$ is the circuit power consumption during the transient
mode period. It is seen from (\ref{total_energy1}) that the active
mode duration $T_{ac}$ is an influential factor in energy saving, as
$E_L$ is a monotonically decreasing function of $T_{ac}$.

\section{In-Body Channel Model}\label{Channel_Model}
An important element in the optimum design of implantable devices is
to characterize a realistic channel model for the human body. For
this purpose, we first study the tissue characteristics and their
effects on the signal propagation loss across the human body over
the MICS frequency band.

Human body is an inhomogeneous and lossy medium which consists of
various tissue layers with different thicknesses and their own unique
electrical characteristics over the frequency of interest. Each
tissue layer absorbs the electromagnetic energy which results in a
considerable attenuation of ultra low power waves propagated from
the IBS to the CCU. In this work, we consider three different tissue layers
composed of muscle, fat and skin, approximating the structure of a
human torso for modeling the channel between the IBS and the CCU.
This model is used in many recent published works (e.g., see
\cite{YazdandoostIEICE0209, ChoITMT1009}).
Table II details the frequency-dependent
permittivity $\varepsilon_r$ and conductivity $\sigma$ of the
aforementioned tissue of an adult male at 403.5 MHz
\cite{YazdandoostIEICE0209, NIREMF2009}. The electrical properties
of the body tissues help in designing proper antennas for implantable
devices, and the in-body path loss channel model. The physical radio propagation with implanted
antennas through human tissues has been extensively studied in the
open literature over the past few years (e.g., see
\cite{SaniITMT1009} and its references), and is out of scope of this
work. Derivation of the exact expression for the in-body channel model is extremely difficult.
\begin{table}
\label{table002} \caption{Electrical Properties of the main tissue layers
of an adult human body at 403.5 MHz} \centering
  \begin{tabular}{lcc}
  \hline
   \textbf{Tissue}           & $\varepsilon_r$   & $\sigma$ (S/m)  \\
  \hline
   Skin (Dry)       & $46.706$             &  $0.6895$             \\
   Fat              & $5.5783$             &  $0.0411$             \\
   Muscle           & $57.10$              &  $0.7972$             \\
  \hline
  \end{tabular}
\end{table}
Reference \cite{Kamran_PIMRC2009} uses a sophisticated 3D virtual
simulation platform to extract a simple statistical path loss model
for the in-body channel over the MICS frequency band. We denote $P_t$
and $P_r$ as the transmitted and the received signal powers for the
uplink, respectively. In addition, it is assumed that the IBS and the CCU are
separated by distance $d$. In this case, the gain factor $\mathcal{L}_d$
for a $\eta^{th}$-power path loss channel is expressed as
\begin{eqnarray}
\mathcal{L}_d & = &
\frac{P_t}{P_r}=\mathcal{L}_{0}\left(\dfrac{d}{d_0}\right)^\eta
\chi_r\\
\mathcal{L}_d (\textrm{dB})&=& \mathcal{L}_{0} (\textrm{dB})+10\eta
\log_{10} \left(\frac{d}{d_0} \right)+\chi_r (\textrm{dB}),
\end{eqnarray}
where $\mathcal{L}_{0}$ is the gain factor at the reference distance
$d_0$ which is specified by the transmitter and the receiver antenna
gains and wavelength $\lambda$, $\eta$ is the path loss exponent, and $\chi_r \sim
N(0,\sigma^2_{_{\chi}})$ is a normal random variable which
represents the deviation caused by different tissue layers (e.g., skin,
muscle, fat) and the antenna gain in different directions
\cite{Kamran_PIMRC2009}. Table III shows the parameters for the
statistical path loss model corresponding to the channel between the
IBS and the CCU for near-surface (e.g., cardiac pacemaker located in
the left pectoral region) and deep tissue (e.g., digestive
endoscopy) applications.

\begin{table}
\label{table003} \caption{Parameters for the path loss model of a
human body at 403.5 MHz \cite{Kamran_PIMRC2009}} \centering
  \begin{tabular}{lccc}
  \hline
   \textbf{Tissue}           & $\mathcal{L}_{0}$ (dB)   & $\eta$   &  $\sigma_{_{\chi}}$ (dB) \\
  Deep Tissue       & $47.14$           & $4.26$   &  $7.85$           \\
  Near Surface      & $49.81$           & $4.22$   &  $6.81$           \\
  \hline
  \end{tabular}
\end{table}

For the above channel model and denoting $s_i(t)$ as the RF
transmitted signal with energy $E_{t}$, the received signal at the
CCU is given by
$r_{i}(t)=\frac{1}{\sqrt{\mathcal{L}_d}}s_{i}(t)+z_{i}(t)$, where
$z_{i}(t)$ is the AWGN at the CCU with two-sided power spectral
density given by $\frac{N_{0}}{2}$. It should be noted that the
power spectral density of the noise introduced by the receiver
front-end at the CCU is calculated by $N_0=\kappa T_0
10^{\frac{NF}{10}}$ (W/Hz), where $\kappa=1.3806503\times 10^{-23}$
is the Boltzmann constant, $T_0$ is the body temperature (in Kelvin)
and NF is the noise figure of the receiver at the CCU. Under the
above considerations, the Signal-to-Noise Ratio (SNR), denoted by
$\gamma$, can be computed as
$\gamma=\frac{E_t}{\mathcal{L}_d N_0}$.

\section{Uncoded FSK Modulation}\label{uncoded_MFSK}
A challenge that arises with implantable medical devices is to
design a small size and ultra low power transceiver which operates
efficiently over the MICS frequency band. With this observation in
mind, finding the energy efficient modulation with low-complexity
implementation is a crucial task in the design of tiny biosensors.
Broadly, the transceivers can be divided into two categories based
on the following modulation techniques: $i)$ sinusoidal
carrier-based and $ii)$ UWB schemes. UWB modulation schemes
such as OOK benefit in very simple design in transmitters along with
low power consumption in circuits. However, according to the FCC regulations for
UWB systems \cite{Oppermann2004}, UWB technology is used solely
for transmitting information over a wide bandwidth at least 500 MHz,
which is much wider than the bandwidth used in the MICS standard. In
addition, the 3-10 GHz frequency band used for UWB systems is far
from the 402-405 MHz frequency range in implantable devices.
On the other hand, sinusoidal carrier-based modulation schemes can
meet the MICS specifications, in particular the frequency band
402-405 MHz and the data rate of at least 250 kbps. Among various
sinusoidal carrier-based modulation techniques, Frequency Shift
Keying (FSK), known as the \emph{green modulation}, has been found to provide a good compromise between
high data rate, simple radio architecture, low power consumption,
and requirements on linearity of the modulation scheme \cite{JamshidICASSP2010, Jamshid_IET2010}. Note that
more complex modulation schemes such as QAM which are often used in
conventional RF communication applications are not easily amenable
to the ultra low power communication demanded by implantable medical devices,
due to higher power consumption.

\begin{table*}
\label{table004} \caption{Parameters for an uncoded NC-MFSK over
AWGN channel with path loss} \centering
  \begin{tabular}{ll}
  \hline
   ~~~~~~~~~~~~~~~~\textbf{Parameter}                     & ~~~~~~~~~~~~~~~~~~~~~~~~~~\textbf{Formula} \\
  \hline
  Transmitted Signal               &  $s_{i}(t)=\sqrt{\frac{2E_t}{T_s}}\cos(2\pi(f_0+i\Delta f)t)$, $i=0,1,...,M-1$   \\
  Symbol Duration                  &  $T_s$ \\
  First Carrier Frequency          &  $f_0$ \\
  Minimum Carrier Separation       &  $\Delta f=\frac{1}{2T_s}$\\
   Channel Bandwidth (Hz)          & $B \approx M\times\Delta f=\frac{M}{2T_s}$    \\
  Data Rate (b/s)                  & $R=\frac{b}{T_s}$ \\
  Bandwidth Efficiency (b/s/Hz)    & $ B_{eff} =  \frac{R}{B}=\frac{2\log_{2}M}{M}$  \\
  Active Mode Duration             & $T_{ac} =  \frac{L}{b}T_s=\frac{ML}{2B\log_2M}$ \\
  Bit Error Rate                   & $P_b \leq \frac{M}{4}e^{-\frac{\gamma}{2}}=\frac{M}{4}e^{-\frac{E_t}{2N_0 \mathcal{L}_d}}$ \\
  Transmit Energy Per Symbol       & $E_t = P_t T_s \approx 2 \mathcal{L}_d N_0 \ln \frac{M}{4P_b}$  \\
  Power Consumption of IBS Circuitry  &   $P_{ct}=P_{Sy}+P_{Filt}+P_{Amp}$\\
  Power Consumption of CCU Circuitry  &  $P_{cr}=P_{LNA}+M\times(P_{Filr}+P_{ED})+P_{IFA}+P_{ADC}$   \\
  \hline
  \end{tabular}
\end{table*}

An FSK scheme is a very simple and low power structure which is
widely utilized in energy-constrained wireless applications
\cite{BohorquezIJSC0409, Jamshid_IET2010} and the IEEE 802.15.6
WPAN Work Group (WG) (e.g., see \cite{IEEE_P802_15}). In an $M$-ary
FSK modulation scheme, $M$ orthogonal carriers are mapped into $b
= \log_{2}M$ bits. The main advantage of this orthogonal
signaling is that the received signals at the CCU do not interfere
with one another in the process of detection at the receiver. An
MFSK modulator benefits in using the Direct
Digital Modulation (DDM) approach, meaning that modulation is
implemented digitally inside the frequency synthesizer. This
property makes MFSK consume very little power and has a faster
start-up time than other sinusoidal
carrier-based modulations. The output of the
frequency synthesizer can be frequency modulated and controlled
simply by $b$ bits in the input of a ``digital control" unit. The
modulated signal is then filtered again, amplified by the Power
Amplifier (PA), and finally transmitted across the human body
channel. We denote the power consumption of the frequency synthesizer,
filters and the power amplifier as $P_{Sy}$, $P_{Filt}$ and $P_{Amp}$,
respectively. In this case, the circuit power consumption of the IBS
is given by $P_{ct}=P_{Sy}+P_{Filt}+P_{Amp}$, where $P_{Amp}=\alpha
P_{t}$ and the value of $\alpha$ is determined based on type of the
power amplifier. Also, denoting $P_{LNA}$, $P_{Filr}$, $P_{ED}$,
$P_{IFA}$ and $P_{ADC}$ as the power consumption of Low-Noise
Amplifier (LNA), filters, envelope detector, IF amplifier and ADC,
respectively, the power consumption of the CCU circuitry with the
Non-Coherent M-ary FSK (NC-MFSK) can be obtained as
$P_{cr}=P_{LNA}+M\times(P_{Filr}+P_{ED})+P_{IFA}+P_{ADC}$
\cite{Jamshid_IET2010}. In addition, it is shown that the energy
consumption during $T_{tr}$ is obtained as $P_{tr}T_{tr}=1.75
P_{Sy}T_{tr}$ \cite{Jamshid_IET2010}.

The energy and bandwidth efficiency of FSK have been extensively
studied in the literature (e.g., see \cite{Jamshid_IET2010,
Cui_GoldsmithITWC0905}). Table IV summarizes the parameters of an
uncoded NC-MFSK over an AWGN channel with path loss which is the
same as the channel model for the human body. There are some
interesting points extracted from Table IV.

$i)$ It is observed that the active mode duration $T_{ac}$ is a
monotonically increasing function of constellation size $M$, when
$L$ and $B$ are fixed. Since, we are interested in having $T_{ac}$ as
small as possible for energy saving (according to
(\ref{total_energy1})), as well as having a low complexity detector
at the CCU, the higher order of constellation size $M$ for the above
NC-MFSK is not desirable. Another advantage of using a lower order
of $M$ is to improve the bandwidth efficiency which is the main
concern in MFSK scheme. Of course, a good spectral
efficiency is achieved using Minimum-Shift Keying (MSK) and in
particular Gaussian Minimum-Shift Keying (GMSK) as a special case of
FSK.

$ii)$ Using $B=\frac{M}{2T_s}=300$ KHz, the data rate is obtained as
$R=\frac{b}{T_s}=2B\frac{\log_2 M}{M}$ which must be greater than
250 kbps according to the MICS specifications in Table I. It is seen
that only $M=2$ and $M=4$ satisfy this requirement emphasizing using
the lower order of $M$ for MFSK modulation scheme.

$iii)$ Assuming $M$ and $L$ are fixed, it is revealed from
$T_{ac} =\frac{ML}{2B\log_2M}$ that the
active mode duration in the IBS with $B=300$ KHz is approximately
20\% of $T_{ac}$ for the wireless sensor applications with the same
NC-MFSK structure, where they use the bandwidth $B=62.5$ KHz according
to Table I. Thus, an 80\% energy saving is achievable for
implantable devices with the MICS standard as compared to the IEEE
802.15.4 physical layer protocol with the same structure used for
sensor networking applications. This interesting result justifies
the reason of using a wider bandwidth in the MICS standard than that
of the IEEE 802.15.4 protocol for WSNs.

Now, we are ready to derive the total energy consumption of an
NC-MFSK based on the results in Table IV. Recall from $E_t
= P_t T_s \approx 2 \mathcal{L}_d N_0 \ln \frac{M}{4P_b}$,
we have
\begin{equation}\label{Power01}
P_t T_{ac}=E_t \frac{T_{ac}}{T_s}= 2 \dfrac{\mathcal{L}_d N_0
L}{\log_2 M} \ln \frac{M}{4P_b}.
\end{equation}
Substituting (\ref{Power01}), $T_{ac}= \frac{ML}{2B\log_2M}$ and
$P_{Amp}=\alpha P_{t}$ in (\ref{total_energy1}), the total energy
consumption of an NC-MFSK scheme for transmitting $L$ bits during
the uplink transmission interval and for a given $P_b$ is obtained
as
\begin{eqnarray}
\notag E_L &=& 2(1+\alpha)  \mathcal{L}_d N_0 \dfrac{L}{\log_2 M}
\ln \frac{M}{4P_b}+\\
\label{energy_totFSK}&&(P_{c}-P_{Amp})\dfrac{ML}{2B\log_{2}M}+1.75
P_{Sy}T_{tr}.
\end{eqnarray}

In transmitting the vital physiological signs of patients, the BER should be
as small as possible. Since, the BER is a function of the received
SNR and to maintain the BER in a specific limit, one would increase
the transmit signal power which is not desirable in ultra low power
implantable medical devices with EIRP=-16 dBm. One practical solution to
avoid an increase in the transmit power is to use channel coding
schemes. For energy optimal designs, however, the impact of channel
coding on the energy efficiency computed in (\ref{energy_totFSK})
must be considered as well. It is a well known fact that channel
coding is a fundamental approach used to improve the link
reliability using redundant information bits along with the
transmitter energy saving due to the providing of coding
gain \cite{Proakis2001}. However, the energy saving comes at the cost of
extra energy consumed in transmitting the redundant bits in codewords
as well as the additional energy consumption in the process of
encoding/decoding. For a certain transmission distance $d$, if these
extra energy consumptions outweigh the transmit energy saving due to
the channel coding, the coded system would not be energy efficient
compared with an uncoded system. In the subsequent sections, we will
argue the above issue for LT codes and show that the LT coded
NC-MFSK surpasses the distance constraint (or equivalently the
dynamic environment between the IBS and the CCU) achieving a given
BER in the proposed system.

\section{Energy Efficiency and Reliability of LT Codes }\label{analysis_Ch4}
The task of the IBS is to communicate vital signs of the patient as
accurately as possible to the CCU, due to the potentially
life-threatening situations in patient monitoring. This is called
data \emph{reliability} which depends strongly on the channel
conditions. Reliability in a wireless network, in general, is
evaluated in terms of the probability of packet loss or the
probability of a data packet being delivered correctly to the
receiver. Enhanced reliability is achieved by adding redundant
bits in data packets in the FEC coding stage at the cost
of an extra power consumption. Thus, there exists a tradeoff between
a higher reliability and a lower power consumption in using FEC
codes. Finding a proper FEC coding scheme for balancing energy
consumption, complexity and data reliability is a challenging task in
designing implant WBANs; in particular, when data transfer across
tissues is susceptible to loss and transmission errors.
The emerging LT codes (as the first practical rateless codes) have
exhibited strong capabilities in reaching the above targets over the
lossy channel model for the human body for the following reasons:

$\textbf{i)}$ \textbf{Optimality:} It is shown in
\cite{LubyFOCS2002} that LT codes are near optimal erasure
correcting codes without the knowledge of the channel erasure rate
at the transmitter.

$\textbf{ii)}$ \textbf{Code-Rate Flexibility:} To ensure reliable
communication across tissues, a sufficient amount of energy must
arrive at the CCU. For fixed-rate codes (e.g., linear block codes),
this may be done by adaptively adjusting the transmitted power based
on the channel condition. Such an adaptive power control is not
feasible for the IBS due to the complexity concerns for implantable
medical devices. LT codes, on the other hand, can adapt to different channel
realizations via changing the code rate (instead of the power
control) to achieve a certain BER. In addition, they have the
capability of high degree of reliability in correct delivery of data
packets over a lossy channel.

$\textbf{iii)}$ \textbf{Simplicity:} The LT encoding process is
extremely simple when compared with some traditional block codes
such as Reed Solomon (RS) codes and Low-Density-Parity-Check (LDPC)
codes, and has a tight power budget. In addition, most of the
complexity of an LT code (i.e., message passing decoder) is pushed
to the CCU which has effectively unlimited power and computational
resource.

The above capabilities of LT codes have removed the critical obstacles of
using the classical fixed-rate codes (e.g., block and convolutional
codes), where the transferred data experience different packet loss
or error rates in different SNRs due to the dynamic channel
conditions for the human body. In this section, we briefly introduce
some basic concepts and definitions for the LT codes. Then, we
analyze the energy efficiency and the reliability of LT coded
NC-MFSK for the proposed implant WBAN. To get more insight into how
LT codes affect the circuit and RF signal energy consumptions in the
system, we modify the energy concepts in Section \ref{uncoded_MFSK},
in particular, the total energy consumption expression in
(\ref{energy_totFSK}) based on the LT coding gain and code rate.

\begin{figure}[t]
\centerline{\psfig{figure=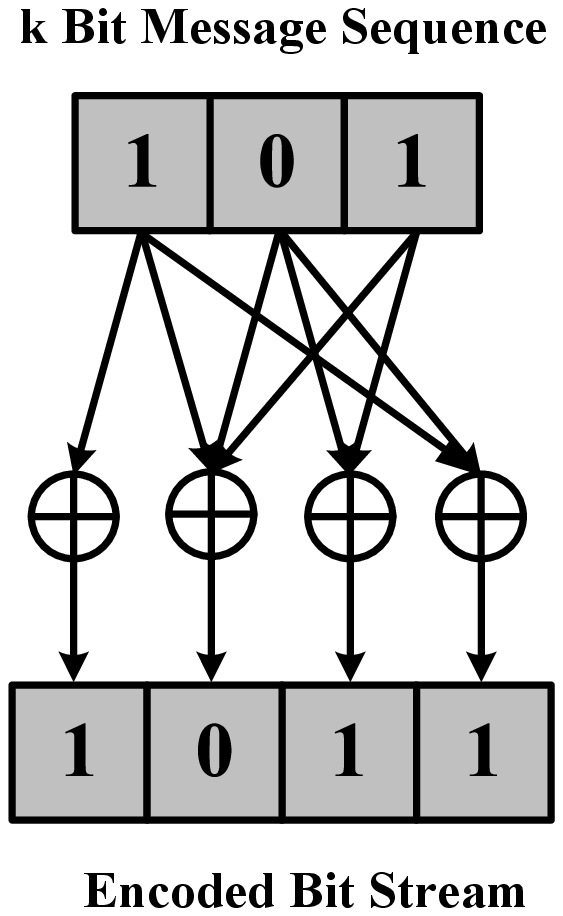,width=1.00in}} \caption{
Factor graph of an LT code with $k=3$ and $n=4$. } \label{fig: LT
Encoder}
\end{figure}

Following the notation of \cite{Shokrollahi_ITIT0606}, an LT code is
specified by the number of input bits $k$ and the output-node degree
distribution $\Omega(x) = \sum_{i=1}^{k}\Omega_i x^i$,
where $\Omega_i$, $i=1,...,k$, denotes the probability that an
output node has degree $i$. Without loss of generality and for ease
of our analysis, we assume that a finite $k$-bit message
$\mathbb{B}_1 = \lbrace{ m_i \rbrace}_{i=1}^{k}\in
\mathbb{M}_L$ is encoded to the codeword $\mathbb{C}_1 =
\lbrace{a_j \rbrace}_{j=1}^{n}$ with $\Omega(x):\lbrace{ m_i
\rbrace}_{i=1}^{k}\rightarrow a_j$, $j=1,...,n$. More precisely,
each single coded bit $a_j$ is generated based on the encoding
protocol proposed in \cite{LubyFOCS2002}: $i)$ randomly choose a
degree $1 \leq \mathcal{D} \leq k$ from a priori known degree
distribution $\Omega(x)$, $ii)$ using a uniform distribution,
randomly choose $\mathcal{D}$ distinct input bits, and calculate the
encoded bit $a_j$ as the XOR-sum of these $\mathcal{D}$ bits. The
above encoding process defines a \emph{factor graph} connecting
encoded nodes to input nodes (see, e.g., Fig. \ref{fig: LT
Encoder}). It is seen that the LT encoding process is extremely
simple and the encoder has a much lower computational burden than
the block codes. For instance, the complexity of the LT codes is of
order $O(n \log n)$ compared to $O(n^2)$ for RS codes used in WiMAX
in the IEEE 802.16 standard. This low complexity comes from
exploiting XOR operations in LT codes (unlike field operations used
for RS codes). Thus, one can with a good approximation assume that
the energy consumption of an LT encoder is negligible compared to
the other circuit components in the IBS. As a result, the energy
cost of the IBS circuitry with the LT coded NC-MFSK scheme is
approximately the same as that of the uncoded one.

One inherent property of LT codes is that they can vary their
codeword block lengths to match a wide range of possible channel
conditions. In fact, for a specific value of SNR, $a_n \in
\mathbb{C}_1$ is the last bit generated at the output of the LT
encoder before receiving the acknowledgement signal from the CCU in
the uplink path indicating termination of a successful decoding
process. This is in contrast to classical linear block and
convolutional codes, in which the codeword block length is fixed. As
a result, the LT code rate, denoted by $R_c = \frac{k}{n}$,
displays a random variable indicating the rateless behavior of the
LT code.

We now turn our attention to the LT code design for the augmentation
protocol. As seen in the LT encoding process, the output-node degree
distribution $\Omega(x)$ is the most influential factor in the
complexity of the LT code. Typically, optimizing $\Omega(x)$ for a
specific wireless channel is a crucial task in designing LT
codes. In this work, we use the following output-node degree
distribution which was optimized for a BSC using a hard-decision
decoder \cite{David_Thesis2008, JamshidQBSC2010}:
\begin{eqnarray}
\notag \Omega(x)&=&0.00466x+0.55545x^2+0.09743x^3+\\
\notag && 0.17506x^5+0.03774x^8+0.08202x^{14}+\\
\label{degree}&& 0.01775x^{33}+0.02989x^{100}.
\end{eqnarray}

The LT decoder at the CCU is able to reconstruct the entire $k$-bit
message $\mathbb{B}_1$ with a high degree of reliability after
receiving any $(1+\epsilon)k$ bits in its buffer, where $\epsilon$
depends upon the LT code design and channel condition
\cite{Shokrollahi_ITIT0606}.
In this work, we assume that the CCU
recovers a $k$-bit message $\mathbb{B}_1$ using a simple hard-decision
``\emph{ternary message passing}" decoder in a nearly identical
manner to the ``\emph{Algorithm E}'' decoder in
\cite{RichardsonITIT0201} for Low-Density Parity-Check (LDPC)
codes. Description of the \emph{ternary message passing}
decoding is out of scope of this work, and the reader is referred to
Chapter 4 in \cite{David_Thesis2008} for more details. It should be
noted that the degree distribution $\Omega(x)$ in (\ref{degree}) was optimized for
a ternary decoder in a BSC and we are aware of no better $\Omega(x)$
for the ternary decoder in AWGN channels.

\begin{figure}[bhpt]
\centerline{\psfig{figure=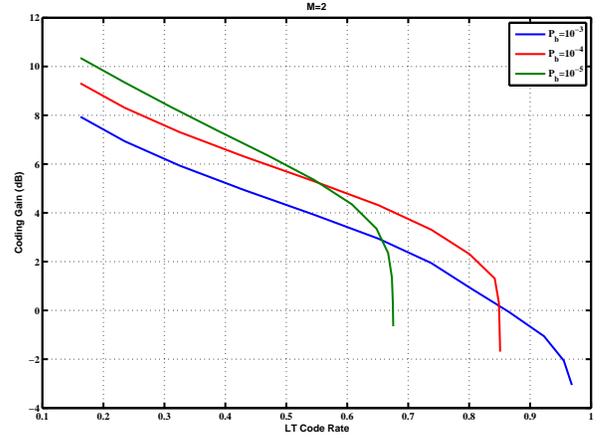,width=3.70in}}
\vspace{-7pt} \center{\hspace{16pt} \small{(a)}} \vspace{10pt}
\hspace{1pt} \centerline{\psfig{figure=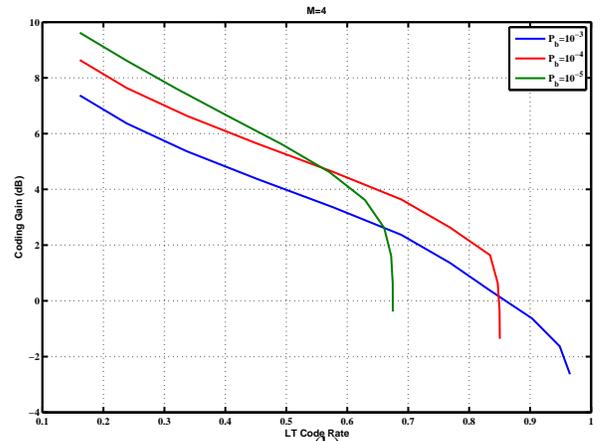,width=3.70in}}
\vspace{-35pt}
\center{\hspace{14pt} \small{(b)}} \\
\vspace{-7pt} \caption[a.] {LT coding gain versus LT code rate for different values of BER and for {a) $M=2$, b) $M=4$ .}}
\label{fig: LT code_rate_coding_gain}
\end{figure}

To analyze the energy efficiency of the LT coded NC-MFSK scheme, we
note that the number of transmitted bits during the uplink
transmission period is increased from the $L$-bit uncoded message to
$\frac{L}{R_c}=\frac{L}{k}n$ bits coded one. In order to keep the
bandwidth of the coded system the same as that of the uncoded case,
we must keep the information transmission rate constant, i.e., the
symbol duration $T_{s}$ of uncoded and coded NC-MFSK would be the
same. According to Table IV, however, the active mode duration
increases from $T_{ac}=\frac{L}{b}T_{s}$ in the uncoded system to
\begin{equation}\label{LT_active}
T_{ac,c}=\frac{L}{bR_{c}}T_{s}=\frac{T_{ac}}{R_{c}}=\dfrac{ML}{2R_cB\log_2M},
\end{equation}
for the LT coded case, where we use $T_{ac}=\frac{ML}{2B
\log_2 M}$ in Table IV. An interesting point raised from
(\ref{LT_active}) is that $T_{ac,c}$ is a function of the random
variable $R_c$ which results in an inherent adaptive duty-cycling
for power management in each channel condition. To compute the total
energy consumption of coded scheme, we use the fact that channel
coding can reduce the required SNR value to achieve a given BER.
Denoting $\gamma =\frac{E_t}{\mathcal{L}_d N_0}$ and
$\gamma_c =\frac{E_{t,c}}{\mathcal{L}_d N_0}$ as the SNR of uncoded
and coded schemes, respectively, the \emph{coding gain}
$\mathcal{G}_{c}$ (expressed in dB) is defined as the difference
between the values of $\gamma$ and $\gamma_c$ required to achieve a
certain BER, where $E_{t,c}=\frac{E_{t}}{\mathcal{G}_{c}}$.
Taking this into account, the proposed implant system with LT codes
benefits in transmission energy saving specified by
$\mathcal{G}_{c}$. Tables V and VI (also see Fig. \ref{fig: LT code_rate_coding_gain}) give the LT code rates and
the corresponding coding gains of LT coded NC-MFSK scheme using
$\Omega(x)$ in (\ref{degree}) and for $M=2$ and 4 and different
values of BER. It is observed that the LT code is able to provide a
large value of coding gain $\mathcal{G}_{c}$ given BER, but this
gain comes at the expense of a very low code rate, which means many
additional code bits need to be sent. This results in higher energy
consumption per information bit. An interesting point extracted from
Fig. \ref{fig: LT code_rate_coding_gain} is the flexibility of the LT code to adjust its rate
(and its corresponding coding gain) to suit channel conditions in
implant WBANs. For instance, in the case of favorable channel
conditions, the LT coded NC-MFSK is able to achieve a code rate
approximately equal to one with $\mathcal{G}_{c}\approx 0$ dB in the
BER=$10^{-3}$, which is similar to the case of uncoded NC-MFSK. The
effect of LT code rate flexibility on the total energy consumption
is also observed in the simulation results in the subsequent
section.

To get more insight into the results of Tables V and VI, we
illustrate the LT code rates versus SNR for several different values
of BER for the cases where $M=2$ and $M=4$ in Fig. \ref{fig:
Rate_SNR}. These results were obtained by evaluating the performance
of the LT code using density evolution, as described in
\cite{David_Thesis2008}. An interesting phenomenon exhibited in the
figure is that the curves tend to converge for a given $M$ as SNR
decreases. This result comes from the fact that for large block
lengths, LT codes exhibit a decoding threshold in much the same way
that LDPC codes do. Essentially, this means that for a particular
SNR there exists a threshold rate above which the LT code will not
successfully decode. At or below this threshold rate, the LT code
\emph{will} decode and the decoded message will have a particular
BER.  As the rate is further decreased, the BER will decrease in a
continuous manner. Thus, there exists a discontinuity in the BER at
the decoding threshold, where the BER of an LT decoded message drops
from being very high (i.e., many errors since the LT decoder failed
to converge to a solution for rates above the threshold) to very low
(i.e., the LT code converged at a successful solution for rates
below the threshold). This discontinuity is more pronounced at lower
SNR, meaning that the BER values of $10^{-3}$, $10^{-4}$, and
$10^{-5}$ all appear to occur at the same points on the rate-SNR
curve.

\begin{table}
\label{table0021} \caption{Code rate and Coding gain($d$B) of LT
coded NC-MFSK over AWGN for different values of $P_b$ and M=2.}
\centering
  \begin{tabular}{clll}
  \hline
    & ~~ $P_b=10^{-3}$ & ~~~$P_b=10^{-4}$  &~~~$P_b=10^{-5}$  \\
   \hline
  SNR (dB) & ~ $R_c$~~~~~~$\mathcal{G}_{c}$ & ~~$R_c$~~~~~~~$\mathcal{G}_{c}$ & ~~$R_c$~~~~~~~$\mathcal{G}_{c}$ \\
   \hline
    3 &  0.1624~~~~7.94  & 0.1624~~~~9.31  & 0.1624~~~~10.35  \\
    4 &  0.2346~~~~6.94  & 0.2346~~~~8.31  & 0.2346~~~~9.35  \\
    5 &  0.3248~~~~5.94  & 0.3248~~~~7.31  & 0.3101~~~~8.35  \\
    6 &  0.4304~~~~4.94  & 0.4304~~~~6.31  & 0.3889~~~~7.35  \\
    7 &  0.5438~~~~3.94  & 0.5438~~~~5.31  & 0.4699~~~~6.35  \\
    8 &  0.6513~~~~2.94  & 0.6513~~~~4.31  & 0.5458~~~~5.35  \\
    9 &  0.7378~~~~1.94  & 0.7378~~~~3.31  & 0.6075~~~~4.35  \\
   10 &  0.8003~~~~0.94  & 0.8003~~~~2.31  & 0.6479~~~~3.35  \\
   11 &  0.8650~~~~-0.06 & 0.8419~~~~1.31  & 0.6671~~~~2.35  \\
   12 &  0.9229~~~~-1.06 & 0.8488~~~~0.31  & 0.6732~~~~1.35  \\
   13 &  0.9553~~~~-2.06 & 0.8498~~~~-0.69 & 0.6748~~~~0.35  \\
   14 &  0.9685~~~~-3.06 & 0.8508~~~~-1.69 & 0.6754~~~~-0.65 \\
    \hline
   \end{tabular}
\end{table}

\begin{table}
\label{table0022} \caption{LT code rate and Coding gain($d$B) of LT
coded NC-MFSK over AWGN for different values of $P_b$ and M=4.}
\centering
  \begin{tabular}{clll}
  \hline
    &  ~~$P_b=10^{-3}$  & ~~~$P_b=10^{-4}$  &~~~$P_b=10^{-5}$  \\
   \hline
  SNR (dB) & ~ $R_c$~~~~~~~~$\mathcal{G}_{c}$&~~$R_c$ ~~~~~~~$\mathcal{G}_{c}$&~~$R_c$~~~~~~~$\mathcal{G}_{c}$ \\
   \hline
    7 & 0.1613~~~~7.37  & 0.1613~~~~8.64  & 0.1613~~~~9.62  \\
    8 & 0.2380~~~~6.37  & 0.2380~~~~7.64  & 0.2380~~~~8.62  \\
    9 & 0.3365~~~~5.37  & 0.3365~~~~6.64  & 0.3190~~~~7.62  \\
   10 & 0.4525~~~~4.37  & 0.4525~~~~5.64  & 0.4051~~~~6.62  \\
   11 & 0.5754~~~~3.37  & 0.5754~~~~4.64  & 0.4925~~~~5.62  \\
   12 & 0.6885~~~~2.37  & 0.6885~~~~3.64  & 0.5709~~~~4.62  \\
   13 & 0.7681~~~~1.37  & 0.7681~~~~2.64  & 0.6289~~~~3.62  \\
   14 & 0.8343~~~~0.37  & 0.8343~~~~1.64  & 0.6605~~~~2.62  \\
   15 & 0.9032~~~~-0.63 & 0.8468~~~~0.64  & 0.6720~~~~1.62  \\
   16 & 0.9486~~~~-1.63 & 0.8498~~~~-0.36 & 0.6746~~~~0.62  \\
   17 & 0.9654~~~~-2.63 & 0.8503~~~~-1.36 & 0.6748~~~~-0.38 \\

   \hline
   \end{tabular}
\end{table}

\begin{figure}[t]
\centerline{\psfig{figure=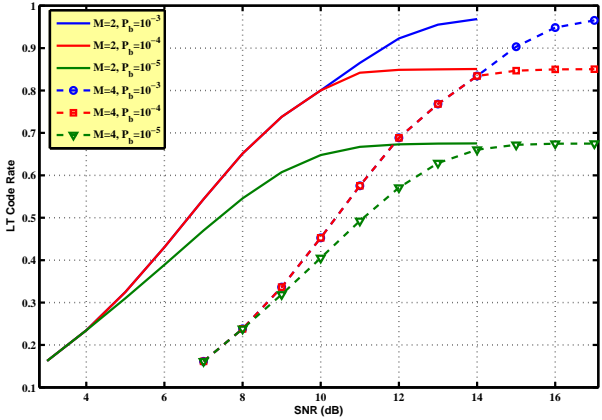,width=4.14in}} \caption{
LT code rate versus SNR for different values of BER and M=2, 4. }
\label{fig: Rate_SNR}
\end{figure}

We denote $E_{enc}$ and $E_{dec}$ as the
computation energy of the encoder and decoder for each information
bit, respectively. Thus, the total computation energy cost of the
coding components for $\frac{L}{R_{c}}$ bits is obtained as
$L\frac{E_{enc}+E_{dec}}{R_{c}}$.
Substituting $E_t = 2
\mathcal{L}_d N_0 \ln \frac{M}{4P_b}$ in
$E_{t,c}=\frac{E_{t}}{\mathcal{G}_{c}}$ and using (\ref{LT_active}),
the total RF signal energy consumption during active mode duration
$T_{ac,c}$ for the coded case is given by
\begin{equation}\label{Power02}
P_{t,c} T_{ac,c}=E_{t,c} \frac{T_{ac,c}}{T_s}= 2
\dfrac{\mathcal{L}_d N_0}{\mathcal{G}_c} \dfrac{L}{R_c\log_2 M} \ln
\frac{M}{4P_b}.
\end{equation}
Substituting (\ref{LT_active}) and (\ref{Power02}) in
(\ref{total_energy1}), and using the fact that $P_{Amp}=\alpha
P_{t,c}$, the total energy consumption of transmitting
$\frac{L}{R_c}$ bits during the uplink transmission period for an LT
coded NC-MFSK scheme, denoted by $E_{L,c}$, and for a given $P_b$ is
obtained as
\begin{eqnarray}
\notag E_{L,c} &=& 2(1+\alpha)  \dfrac{\mathcal{L}_d
N_0}{\mathcal{G}_{c}}\dfrac{L}{R_c \log_2 M} \ln \frac{M}{4P_b}+\\
\notag &&(P_{c}-P_{Amp}) \dfrac{ML}{2R_c B\log_{2}M}+1.75 P_{Sy}T_{tr}+\\
\label{energy_totalcoded1}&& L\frac{E_{enc}+E_{dec}}{R_{c}},
\end{eqnarray}
where the goal is to minimize $E_{L,c}$ in each distance $d$, in
terms of modulation and coding parameters. Although, higher
reliability (corresponding to the lower BER) results in an increase
in the total energy consumption $E_{L,c}$, the effect of the coding
gain $\mathcal{G}_{c}$ and the code rate $R_c$ on $E_{L,c}$ for a
given $P_b$ should be considered as well.

\section{Numerical Evaluation}\label{simulation_Ch5}
This section presents some numerical evaluations using realistic
parameters from the MICS standard and state-of-the art technology to
confirm the reliability and energy efficiency analysis of uncoded
and LT coded NC-MFSK schemes discussed in Sections
\ref{uncoded_MFSK} and \ref{analysis_Ch4}. We assume that the
NC-MFSK modulation scheme operates in the carrier frequency
$f_0=$403.5 MHz according to the frequency band requirement in the
MICS standard. As shown in Section \ref{uncoded_MFSK}, we only
consider the constellation sizes $M=2,4$. We assume that in each
period $T_L$, the sensed data frame size $L=1024$ bytes (or
equivalently $L=8192$ bits) is generated, where $T_L$ is assumed to
be 1.4 seconds. The channel bandwidth is assumed to be
$B=\frac{M}{2T_s}=300$ KHz, according to Table I. As a result, the
symbol duration $T_s$ for $M=2$ and $M=4$ is obtained as $3.33$ $\mu
s$ and $6.67$ $\mu s$, respectively. Also, the data rate
$R=\frac{b}{T_s}$ for both $M=2,4$ will be 300 kbps which satisfies
the data rate requirement in Table I. Assuming $T_0=310^{\circ K}$
for the body temperature and NF=8 dB, the total noise power at the
bandwidth $B=300$ KHz is obtained as $N_0=\kappa B T_0
10^{NF/10}=-110.91$ dBm. Table VII summarizes the system parameters
for simulation. The results in Tables III-VI are also used to
compare the energy efficiency of uncoded and LT coded NC-MFSK
schemes. In addition, the current simulation is based on the values
in Table III for the two following scenarios: $i)$ near surface with the
distance range of $d_0=$30 mm to 100 mm and $ii)$ deep tissue where
the distance $d$ varies between 100 mm and 300 mm.
For this purpose, it is assumed that the skin varies in thickness from
$0.5$ mm to $6$ mm \cite{Wysocki2006} and the muscle thickness varies in
the range of 13-40 mm according to the MRI measurements
in \cite{Dupnot2001}. Furthermore, we assume that the fat thickness
ranges from 0-250 mm for thin or obese patients.
In order to estimate the computation energy of the channel
coding, we use the ARM7TDMI core which is the industry's most widely
used 32-bit embedded RISC microprocessor for an accurate power
simulation \cite{ARMTech2004}.

\begin{table}
\label{table006} \caption{System Evaluation Parameters} \centering
  \begin{tabular}{lll}
  \hline
   $B=300$ KHz         & $T_{tr}=5~\mu s$  & $P_{ADC}=7$ mw \\

   $d_0=30$ mm         & $P_{Sy}=10$ mw    & $P_{LNA}=9$ mw \\

   $N_0=-110.91$ dBm   & $P_{Filt}=2.5$ mw & $P_{ED}=3$ mw  \\

   $T_L=1.4$ sec       & $P_{Filr}=2.5$ mw & $P_{IFA}=3$ mw \\
  \hline
  \end{tabular}
\end{table}

\begin{figure}[bhpt]
\centerline{\psfig{figure=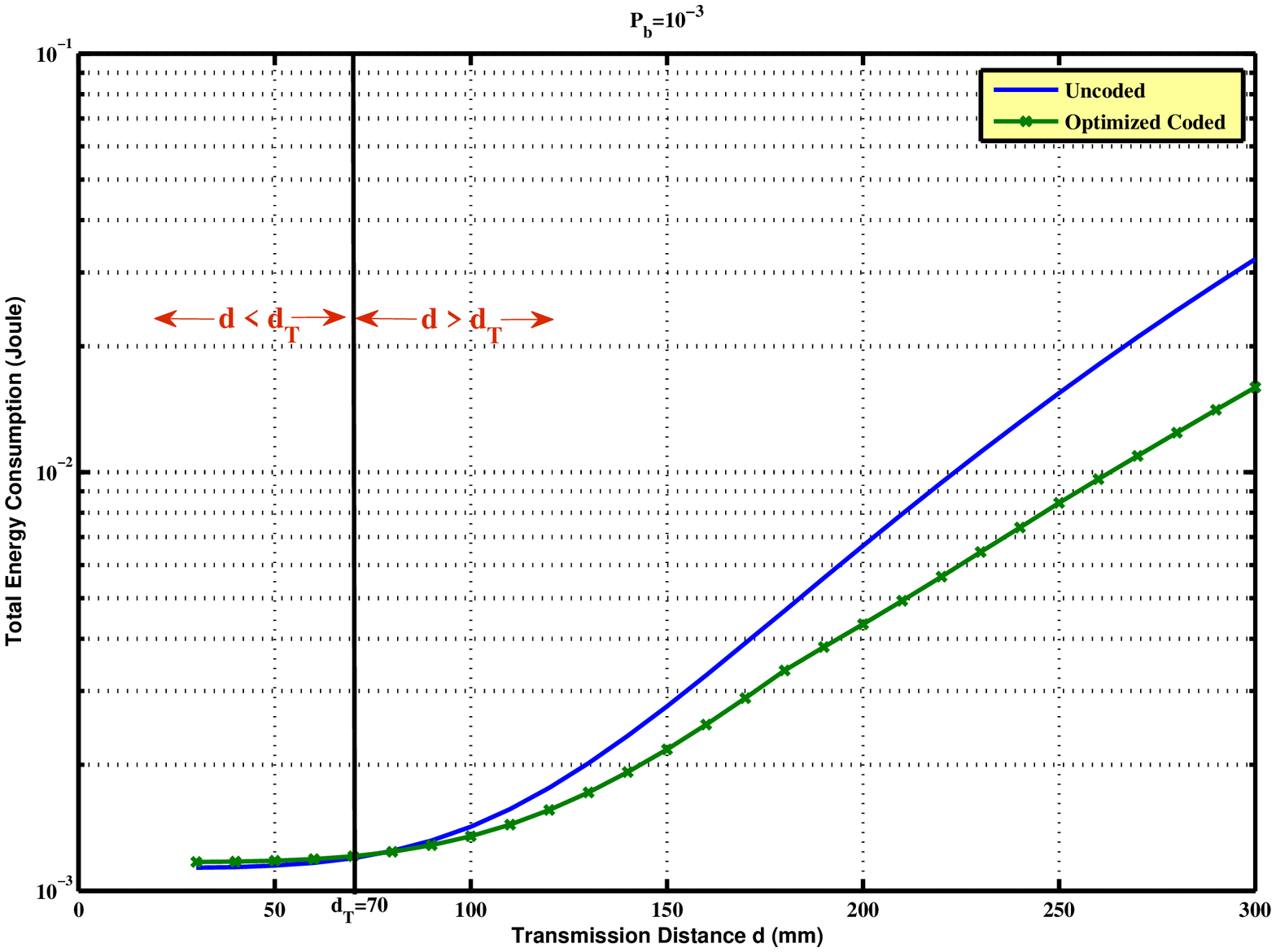,width=3.86in}}
\vspace{-7pt} \center{\hspace{16pt} \small{(a)}} \vspace{10pt}
\hspace{1pt}
\centerline{\psfig{figure=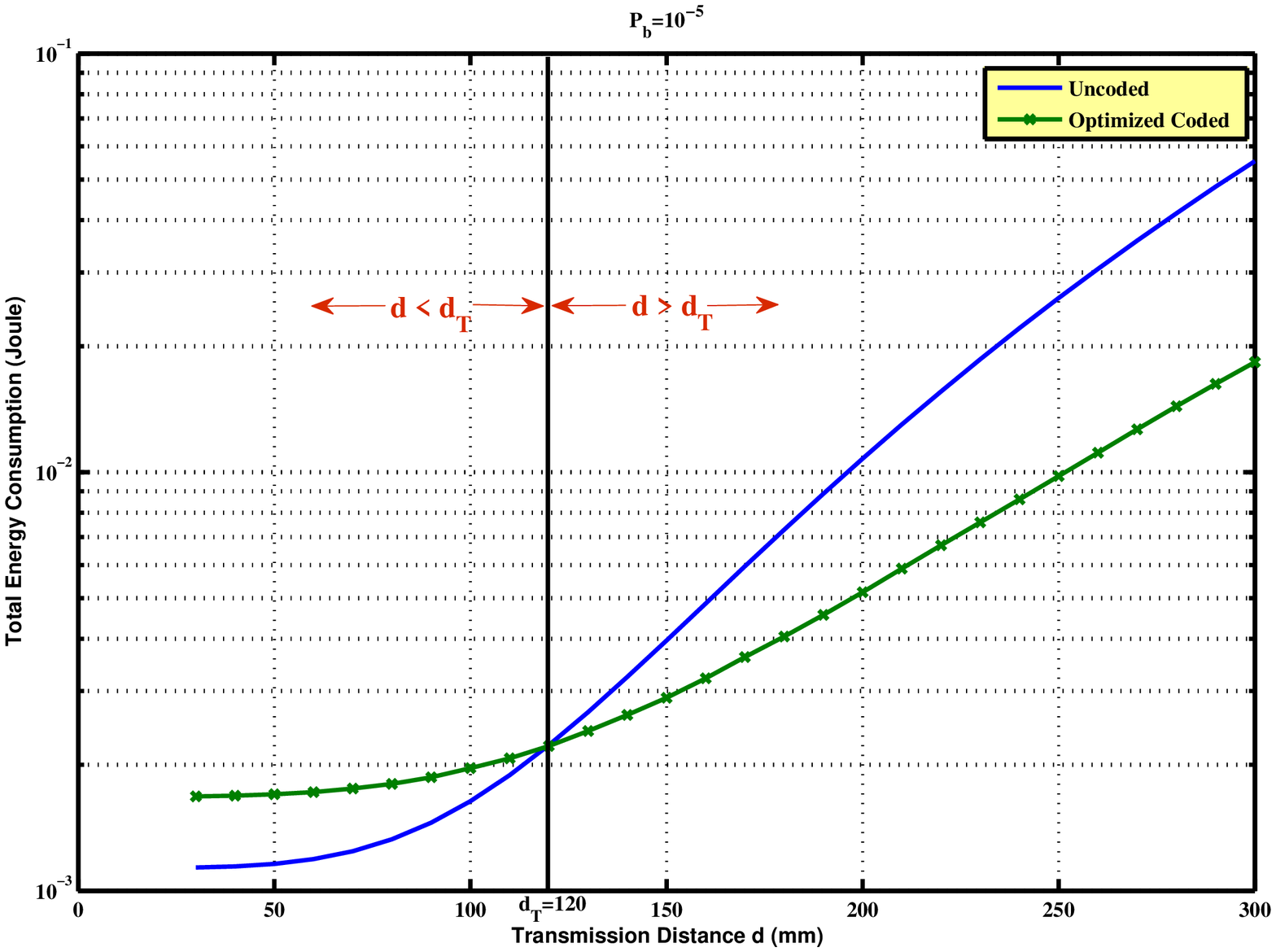,width=3.86in}}
\vspace{-35pt}
\center{\hspace{14pt} \small{(b)}} \\
\vspace{-7pt} \caption[a) $\hat{\alpha}=W=1$.] {\small{Total energy
consumption of uncoded and optimized LT coded NC-MFSK versus $d$ for
a) $P_b=10^{-3}$ b) $P_b=10^{-5}$.}} \label{fig: Coded_Uncoded}
\end{figure}

Fig. \ref{fig: Coded_Uncoded} shows the total energy consumption
versus distance $d$ for the optimized LT coded NC-MFSK compared to
the optimized uncoded NC-MFSK for the cases $P_b=10^{-3}$ and
$P_b=10^{-5}$, and for the proposed three different tissue layers
with the aforementioned thicknesses. The optimization is performed over $M$ and the
parameters of coding scheme in Tables V and VI. For the case of
$P_b=10^{-3}$, it is revealed from Fig. \ref{fig: Coded_Uncoded}-a
that for distance $d$ less than the threshold level $d_T \approx 70$
mm (near surface scenario), the total energy consumption of optimized uncoded NC-MFSK is
less than that of the coded NC-MFSK schemes. However, the energy gap
between the LT coded and uncoded approaches is negligible as
expected. For $d>d_T$ which covers the deep tissue scenario, the LT
coded NC-MFSK scheme is more energy efficient than the uncoded one,
in particular when $d$ grows. This result comes from the high coding
gain capability of LT codes. An interesting phenomenon exhibited in
Fig. \ref{fig: Coded_Uncoded}-a is that the optimized LT coded
NC-MFSK surpasses the distance constraint in implant WBAN
applications. In addition, the flexibility of the LT code rate (and
the corresponding LT coding gain) allows the system to adjust its
duty-cycling for power management and to suit to different channel
conditions for the human body for any distance $d$.

In the case of $P_b=10^{-5}$ as depicted in Fig. \ref{fig:
Coded_Uncoded}-b, the uncoded NC-MFSK scheme is preferable to use
for $d < d_T \approx 120 $ mm which corresponds to the
near-surface applications. This result follows from the maximum code
rate of $R_c \approx 0.675$ for both $M=2$ and $M=4$ in
$P_b=10^{-5}$. However, for the deep tissue applications where $d
> d_T \approx 120$ mm, the optimized LT coded NC-MFSK is much more
energy efficient than the uncoded case.

It is concluded from the above observations that the threshold level
$d_T$ for a certain BER is a fundamental parameter in the physical
layer design of implant WBANs, as it determines \emph{``when the LT
codes are energy efficient''} in the proposed wireless biomedical
implant system. The parameter $d_T$ is obtained when the total
energy consumptions of coded and uncoded systems become equal. Using
$\mathcal{L}_d =\mathcal{L}_{0}\left(\frac{d}{d_0}\right)^\eta
\chi_r$, and the equality between (\ref{energy_totFSK}) and
(\ref{energy_totalcoded1}) for uncoded and LT coded NC-MFSK, we have
\begin{equation}\nonumber
d_T=d_0\left[\dfrac{M
(\mathcal{P}_{c}-\mathcal{P}_{Amp})}{4(1+\alpha)N_0B \mathcal{L}_0
\chi_r \ln \left(\frac{M}{4P_b}
\right)}\dfrac{\mathcal{G}_{c}(1-R_c)}{\mathcal{G}_{c}R_c-1}\right]^{\dfrac{1}{\eta}}.
\end{equation}
It can be seen from the above equation that a decrease in the $P_b$
results in an increase in the $d_T$ as observed in Fig. \ref{fig:
Coded_Uncoded}. In addition, the growth rate of the energy
consumption of the LT codes for deep tissue applications (i.e.,
$d>d_T$) is much smaller than that of the uncoded one when $P_b$
decreases (equivalent to the higher reliability).

\textbf{Remark:} As discussed previously, the proposed LT coded scheme benefits
in adjusting the coding parameters in each channel realization or equivalently
each distance $d$ to minimize the total energy consumption. For this case, as long
as the CCU collects a sufficient number of bits, it will be able to decode
the original message. Indeed, only the CCU needs to be aware of the channel
conditions. The IBS, as the transmitter, always sends an LT encoded stream using constant
modulation and power, and the receiver alone determines how many bits to collect,
their reliability, and ultimately decodes the message. Thus, the transmitter does
not need to change coding rate, because the rate is determined entirely by how many
bits it collects based on the channel quality. There is no channel information
required at the transmitter which reduces the complexity and cost in implementation.

In contrast to the ARQ scheme, the only feedback in the LT coded system is a
single message from transmitter to receiver (at the end of a long block)
indicating that a sufficient number of bits have been collected and it is
time to move on to the next data block. This is not an ARQ scheme, but it is
more similar to an incremental redundancy scheme whereby a message is encoded
using a rate-compatible punctured code.

In addition, a nice feature about the optimized LT code as opposed to the
punctured codes is that there is truly no ``minimum rate'' at which the code
can operate. More precisely, we can choose a code that allows us to drop the
rate as low as we need. With punctured LDPC, it is reported in \cite{Soljanin2006}
that depending upon the operating range, rate, etc. the LDPC
performance drops off dramatically beyond a certain rate threshold--a phenomenon
not observed by the rateless codes.

\section{Conclusion}\label{conclusion_Ch6}
In this paper, we presented an augmentation protocol for the
physical layer of the MICS standard with focus on the energy
efficiency and the reliability of LT coded NC-MFSK scheme over the
MICS frequency band. It was shown that the proposed scheme provides
an inherent adaptive duty-cycling for power management which comes
from the flexibility of the LT code rate. Analytical results
demonstrated that an 80\% energy saving is achievable with the
proposed protocol when compared to the IEEE 802.15.4 physical layer
protocol with the same structure used for wireless sensor networks.
Numerical results have shown a lower energy consumption keeping at
the same time a higher degree of reliability in the LT coded NC-MFSK
scheme for deep tissue scenarios when compared to the uncoded case.
In addition, it was shown that the optimized LT coded NC-MFSK is
capable to surpass the distance constraint in implantable medical devices
for the BER=$10^{-3}$. In fact, since the efficiency of implantable
medical devices is strongly related to the human body structure (e.g.,
obese or thin and tissue composition), the
proposed scheme overcomes the patient's vital information loss and
errors, and operates in a high degree of accuracy over any dynamic
channel conditions for the human body. The introduced protocol is
unique in implant WBAN applications, as it has been explicitly
designed with simplicity and flexibility in mind, and should be
usable in current wireless medical networking hardware.

\section*{Acknowledgment}
The authors would like to thank Ali Tawfiq (University of
Toronto) for graciously editing this paper.


\begin{biography}[{\includegraphics[width=1in,height=1.25in,clip,keepaspectratio]{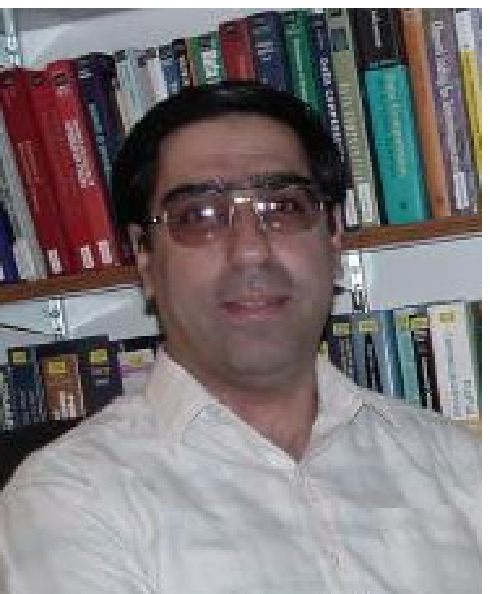}}]
{Jamshid Abouei} received the B.Sc. degree in electronics engineering and
the M.Sc. degree in communication systems engineering (with the highest
honor) both from the Isfahan University of Technology (IUT), Iran, in 1993
and 1996, respectively, and the Ph.D. degree in electrical engineering from
the University of Waterloo in Waterloo, ON, Canada, in 2009.

From 1996 to 2004, he was a faculty member (lecturer) in the Department of
Electrical Engineering, Yazd University, and from 1998 to 2004, he was a
technical advisor and design engineer (part-time) in R\&D center and cable
design department in SGCC company. From 2009 to 2010, he was a Postdoctoral
Fellow in the Multimedia Lab, in the Department of Electrical \& Computer
Engineering, at the University of Toronto, ON, Canada.

Currently, Dr Abouei is an Assistant Professor in the Department of Electrical
\& Computer Engineering, at the Yazd University, Iran. His research interests
are in general areas of wireless ad hoc and sensor networks, with particular
reference to energy efficiency and optimal resource allocation, multi-user
information theory, cooperative communication in wireless relay networks,
applications of game theory, and orthogonal
codes in CDMA systems.

Dr Abouei has received numerous awards and scholarships, including FOE and IGSA
awards for excellence in research in University of Waterloo, Canada, and MSRT Ph.D.
Scholarship from the Ministry of Science, Research and Technology, Iran in 2004.
\end{biography}

\begin{biography}[{\includegraphics[width=1in,height=1.25in,clip,keepaspectratio]{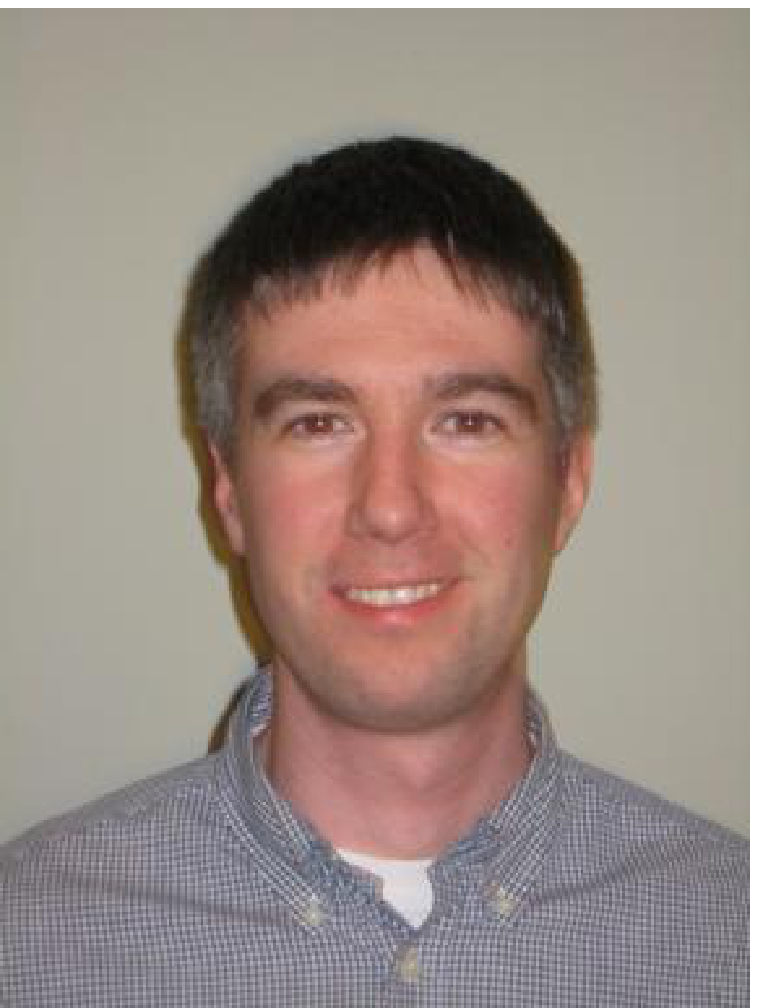}}]
{J. David Brown} was born in Ottawa, ON, Canada, in 1977.  He received the B.Sc.(Eng.)
degree in electrical and computer engineering in 2000, and the M.Sc.(Eng.) degree in
2002, both from Queen's University in Kingston, ON, Canada.  In 2008, he received the
Ph.D. degree in electrical and computer engineering from the University of Toronto in
Toronto, ON, Canada.

From 2002 to 2004 and again from 2007 to 2009, he worked as an Electrical Engineer at
General Motors.  In 2009, he joined the Network Information Operations Section at DRDC
in Ottawa, ON, Canada, as a Research Scientist.  His research interests include digital
communications, error-control codes, and machine learning.

Dr. Brown has received numerous awards and scholarships, including the Natural Sciences
and Engineering Research Council of Canada (NSERC) Post-graduate Scholarship, the NSERC
Canada Graduate Scholarship (CGS), and two Industry Canada Fessenden Postgraduate Scholarships.
He also received the Queen's University Professional Engineers of Ontario Gold Medal.
\end{biography}

\begin{biography}[{\includegraphics[width=1in,height=1.25in,clip,keepaspectratio]{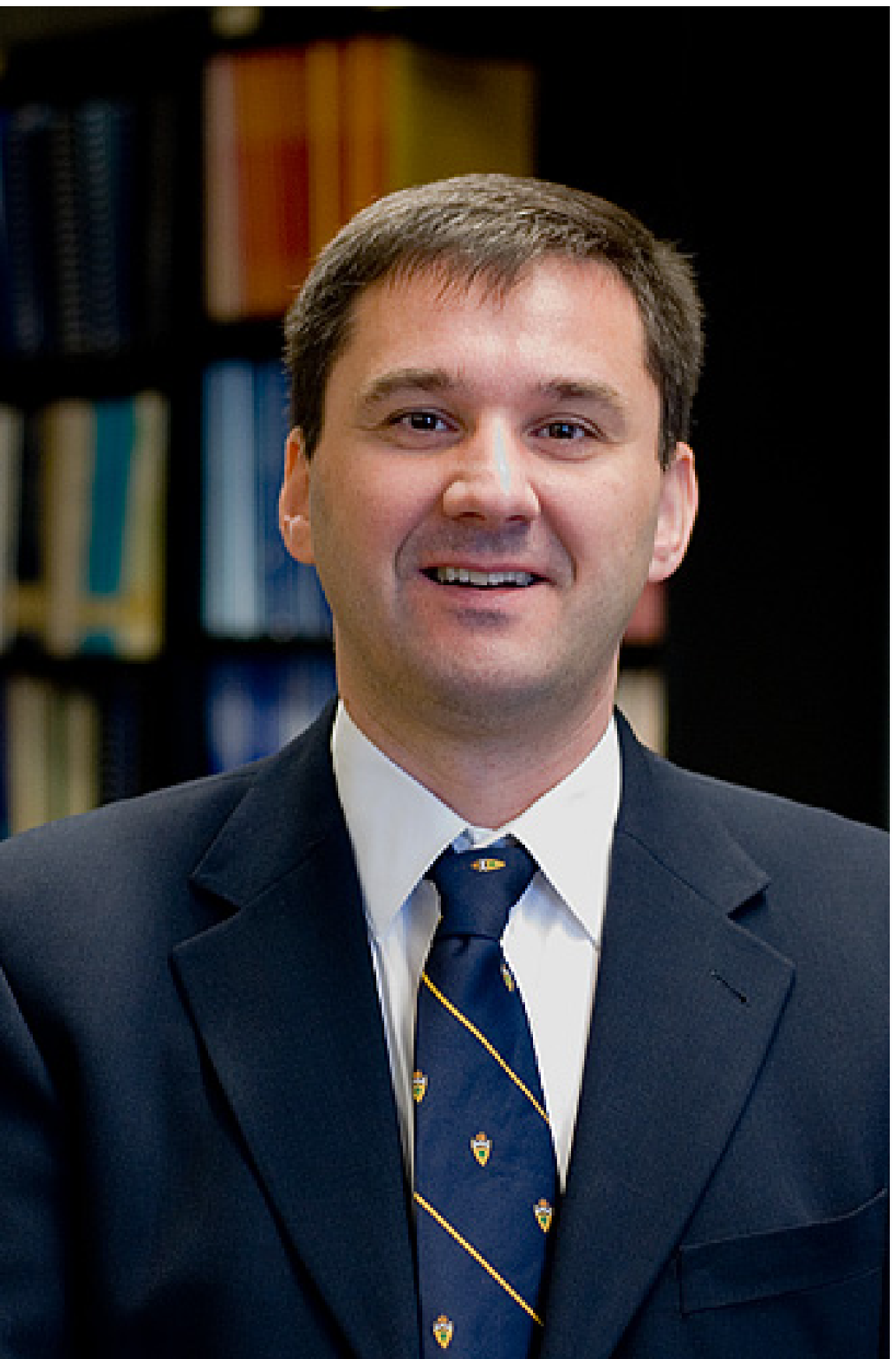}}]
{Konstantinos N. (Kostas) Plataniotis} is a Professor with the
Edward S. Rogers Sr. Department of Electrical and Computer
Engineering at the University of Toronto in Toronto, Ontario,
Canada, and an Adjunct Professor with the School of Computer
Science at Ryerson University, Canada. He is the Director of The
University of Toronto's Knowledge Media Design Institute
(www.kmdi.utoronto.ca), and the Director of Research for the
Identity, Privacy and Security Institute at the University of
Toronto (www.ipsi.utoronto.ca).

Prof. Plataniotis is the Editor in Chief (2009-2011) for the IEEE
Signal Processing Letters and chairs the Examination Committee for
the IEEE Certified Biometrics Professional (CBP) Program
(www.ieeebiometricscertification.org). He served on the IEEE
Educational Activities Board (EAB) and he
was the Chair (2008-09) of the IEEE EAB Continuing Professional
Education Committee. Dr. Plataniotis has served as Chair
(2000-2002) IEEE Toronto Signal Processing Chapter, Chair
(2004-2005) IEEE Toronto Section, and he was a member of the 2006
and 2007 IEEE Admissions \& Advancement Committees.

He is the 2005 recipient of IEEE Canada's Outstanding Engineering Educator Award
``for contributions to engineering education and inspirational guidance of graduate students'' and the co-recipient of the
2006 IEEE Trans. on Neural Networks Outstanding Paper Award for the published in 2003 paper
entitled `` Face Recognition Using Kernel Direct Discriminant Analysis Algorithms''.

He is a registered professional engineer in the province of
Ontario, and a member of the Technical Chamber of Greece, and a Fellow of the Engineering Institute of Canada.

His research interests include biometrics, communications systems,
multimedia systems, and signal \& image processing.

\end{biography}

\begin{biography}[{\includegraphics[width=1in,height=1.25in,clip,keepaspectratio]{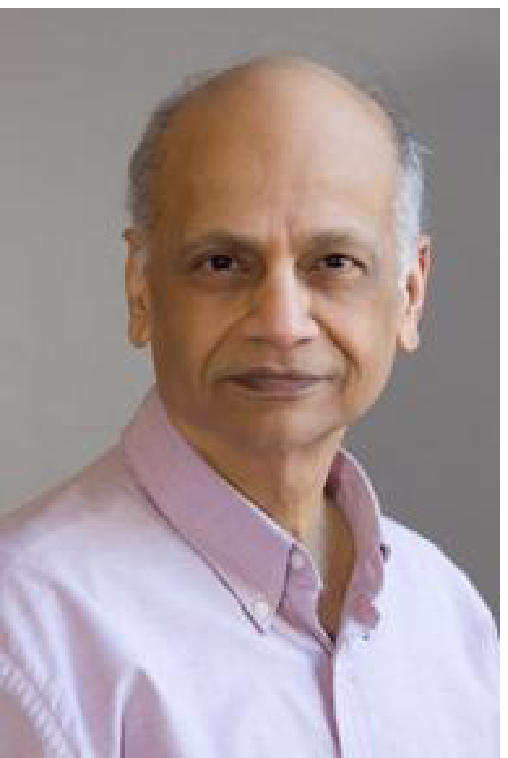}}]
{Subbarayan Pasupathy} was born in Chennai (Madras), Tamilnadu,
India. He received the B.E. degree in telecommunications from the
University of Madras, the M.Tech. degree in electrical engineering
from the Indian Institute of Technology, Madras, and the M.Phil. and
Ph.D. degree in engineering and applied science from Yale University.

Currently, he is a Professor Emeritus in the Department of Electrical
and Computer Engineering at the University of Toronto, where he has
been a Faculty member from 1972. His research over the last three
decades has mainly been in statistical communication theory and
signal processing and their applications to digital communications.
He has served as the Chairman of the Communications Group and as the
Associate Chairman of the Department of Electrical Engineering at the
University of Toronto. He is a registered Professional Engineer in the
province of Ontario. During 1982-1989 he was an Editor for {\it Data
Communications and Modulation} for the IEEE TRANSACTIONS ON COMMUNICATIONS.
He has also served as a Technical Associate Editor for the IEEE COMMUNICATIONS
MAGAZINE (1979-1982) and as an Associate Editor for the {\it Canadian
Electrical Engineering Journal} (1980-1983). He wrote a regular humour column
entitled ``Light Traffic'' for the IEEE COMMUNICATIONS MAGAZINE during 1984-98.

Dr. S. Pasupathy was elected as a Fellow of the IEEE in 1991 ``for contributions
to bandwidth efficient coding and modulation schemes in digital communication'',
was awarded the Canadian Award in Telecommunications in 2003 by the {\it
Canadian Society of Information Theory}, was elected as a Fellow
of the {\it Engineering Institute of Canada} in 2004 and as a Fellow of the
{\it Canadian Academy of Engineering} in 2007. He was honoured as a
Distinguished Alumnus by I.I.T, Madras, India in 2010. He has been
identified as a ``highly cited researcher'' by ISI Web of Knowledge and
his name is listed in ISIHighlyCited.com.
\end{biography}

\end{document}